\documentclass[pdflatex,sn-mathphys-num]{sn-jnl}% Math and Physical Sciences Numbered Reference Style

\usepackage{comment}
\usepackage{graphicx}%
\usepackage{multirow}%
\usepackage{amsmath,amssymb,amsfonts}%
\usepackage{amsthm}%
\usepackage{mathrsfs}%
\usepackage[title]{appendix}%
\usepackage{xcolor}%
\usepackage{textcomp}%
\usepackage{manyfoot}%
\usepackage{booktabs}%
\usepackage{algorithm}%
\usepackage{algorithmicx}%
\usepackage{algpseudocode}%
\usepackage{listings}%
\theoremstyle{thmstyleone}%
\theoremstyle{thmstyletwo}%

\theoremstyle{thmstylethree}%

\begin{document}

\title[Two nonlinearities in a monopolar plasma vortex]{Measuring the tug-of-war between two nonlinearities in a monopolar plasma vortex}

\author[1]{\fnm{Huiwon} \sur{Chung}}\email{hw4450@snu.ac.kr}

\author[2]{\fnm{Gyungjin} \sur{Choi}}\email{gyungjinc@kaist.ac.kr}

\author[3]{\fnm{Donato} \sur{Di Matteo}}\email{dimatteo@riam.kyushu-u.ac.jp}

\author[4]{\fnm{Takuma} \sur{Yamada}}\email{takuma@artsci.kyushu-u.ac.jp}

\author[1]{\fnm{Chweeho} \sur{Heo}}\email{hcho201@snu.ac.kr}

\author[5]{\fnm{Peter} \sur{Manz}}\email{peter.manz@uni-greifswald.de}

\author*[6]{\fnm{Chanho} \sur{Moon}}\email{moon@riam.kyushu-u.ac.jp}

\author*[1]{\fnm{Yong-Su} \sur{Na}}\email{ysna@snu.ac.kr}

\affil[1]{\orgdiv{Department of Nuclear Engineering}, \orgname{Seoul National University}, \orgaddress{\street{1 Gwanak-ro, Gwanak-gu}, \city{Seoul}, \postcode{08826}, \country{Republic of Korea}}}

\affil[2]{\orgdiv{Department of Nuclear and Quantum Engineering}, \orgname{KAIST}, \orgaddress{\street{291 Daehak-ro, Yuseong-gu}, \city{Daejeon}, \postcode{34141}, \country{Republic of Korea}}}

\affil[3]{\orgdiv{Interdisciplinary Graduate School of Engineering Sciences}, \orgname{Kyushu University}, \orgaddress{\street{6-1 Kasuga-koen, Kasuga}, \city{Fukuoka}, \postcode{816-8580}, \country{Japan}}}

\affil[4]{\orgdiv{Faculty of Arts and Science Division for Experimental Natural Science}, \orgname{Kyushu University}, \orgaddress{\street{744 Motooka, Nishi-ku}, \city{Fukuoka}, \postcode{819-0395}, \country{Japan}}}

\affil[5]{\orgname{University of Greifswald}, \orgdiv{Institute of Physics}, \orgaddress{\street{Felix-Hausdorff-Straße 6}, \city{Greifswald}, \postcode{17489}, \state{Mecklenburg-Vorpommern}, \country{Germany}}}

\affil[6]{\orgdiv{Research Institute for Applied Mechanics}, \orgname{Kyushu University}, \orgaddress{\street{6-1 Kasuga-koen, Kasuga}, \city{Fukuoka}, \postcode{816-8580}, \country{Japan}}}

\abstract{
From planetary atmospheres to magnetized plasmas, single-core vortices organize flows and transport heat and matter.
Understanding their persistence requires identifying the nonlinear interactions that sustain a single core.
The Charney--Hasegawa--Mima equation, shared by drift waves and Rossby waves, admits steadily propagating localized vortex solutions only in dipole form.
Theories of monopolar drift vortices attribute the additional nonlinearity to either the electron-temperature gradient or density-profile curvature, but their respective roles remain experimentally unresolved.
Here we show that the coupling sustaining a monopolar drift vortex results from the near cancellation of two much larger nonlinear contributions.
We measure three-wave coupling among the vortex harmonics using simultaneous probe signals acquired around the full azimuth of a laboratory plasma and compare it with predictions of an extended drift-wave equation derived for the measured equilibrium conditions.
The equation reproduces the sign, magnitude and triad dependence of the observed coupling without adjustable parameters, whereas neither contribution alone accounts for the measured coupling pattern.
This coupling transfers fluctuation power from the linearly driven fundamental to damped harmonics, maintaining the single-core waveform.
These results establish an experimental route to distinguishing mechanisms that produce similar vortex structures by resolving the nonlinear interactions within them.
}

\keywords{
Plasma turbulence,
Drift waves,
Coherent structures,
Monopolar vortices,
Nonlinear wave interactions}

\maketitle
\setcounter{secnumdepth}{0}

\par
From Jupiter's Great Red Spot~\cite{marcus1993} and the cyclones encircling its poles~\cite{adriani2018} to eddies in Earth's oceans~\cite{dong2014}, the most persistent vortices in quasi-two-dimensional turbulence typically have a single core.
Such monopolar vortices emerge spontaneously from turbulence and, once formed, dominate its long-time evolution~\cite{mcwilliams1984,xia2011}.
A single core carries a net anomaly of the fluid it traps, so monopolar vortices transport heat and matter as they move, as oceanic eddies do with heat and salt~\cite{dong2014,wang2023}.
In magnetized plasmas, coherent vortices likewise mediate the cross-field transport of particles and heat~\cite{huld1990,nielsen1996,kobayashi2015}.
Understanding what organizes a single-core vortex therefore matters for transport in both geophysical fluids and magnetized plasmas.

\par
Drift waves in magnetized plasmas and Rossby waves in rotating fluids share the potential-vorticity structure of the Charney--Hasegawa--Mima equation~\cite{charney1948,hasegawa1978,hasegawa1979}, whose advective Jacobian nonlinearity admits steadily propagating localized solutions only in dipolar form~\cite{larichev1976,makino1981} (Fig.~\ref{fig:1}a).
A long-lived monopolar vortex therefore requires physics beyond this equation.
In planetary flows, monopolar vortices have been modelled both as vortices stabilized by ambient zonal shear~\cite{marcus1988} and as Rossby solitons sustained by a KdV-type scalar nonlinearity~\cite{petviashvili1980,antipov1986}.
In the drift-wave description, a monopole likewise requires a scalar nonlinearity that breaks the polarity symmetry of the equation, together with a propagation velocity outside the continuum of linear waves, without which the core radiates into its exterior~\cite{petviashvili1976,nycander1989exist,su1991}.
This nonlinearity must be supplied by inhomogeneity of the plasma equilibrium, and the reduced drift-wave equations that admit monopolar vortices attribute it either to the electron-temperature gradient~\cite{petviashvili1980,nycander1989temp} or to the curvature of the density profile~\cite{lakhin1987,lakhin1988}, depending on the asymptotic ordering adopted~\cite{laedkespatschek1986,laedke1988,spatschek1990,nycander1991comment,salat1990,su1991}.
Because both attributions admit monopolar solutions, observing a monopole cannot decide between them, whereas they predict distinct patterns of nonlinear interaction among the harmonics of the vortex.
Yet, whichever ordering is adopted, none of these equations has been tested against the nonlinear interactions measured within a vortex.

\par
In planetary atmospheres, the energy feeding long-lived vortices can now be traced across scales~\cite{siegelman2022}, but such observations cannot isolate the individual nonlinear interactions that the governing equation predicts.
Laboratory experiments, in which both the equilibrium and the fluctuations are accessible, have characterized coherent vortices in magnetized plasmas~\cite{huld1991,grulke1999,arakawa2016,arakawa2022} and rotating fluids~\cite{sommeria1988,zhang2015,lemasquerier2020}, including their morphology, persistence and coupling to surrounding fluctuations.
In linear magnetized plasmas, bispectral measurements have further shown that coherent structures can form through nonlinear coupling among drift waves~\cite{yamada2008}.
Coupling coefficients can also be inferred directly from fluctuation data~\cite{ritz1989,kimjs1996,dewit1999}.
Within a coherent vortex whose harmonics are measured simultaneously around the full azimuth, this inference turns the nonlinear terms of the governing equation into measurable quantities, which can be compared with theory triad by triad.

\par
Here, for the first time, we directly compare drift-wave theory with nonlinear coupling measured within a monopolar drift vortex in the Plasma Assembly for Nonlinear Turbulence Analysis (PANTA), a linear plasma device~\cite{oldenburger2012}.
We derive an extended drift-wave equation that retains both candidate nonlinearities, with coefficients evaluated from the measured equilibrium profiles, and compare its predictions with three-wave coupling measured among the vortex harmonics.
The equation reproduces the sign, magnitude and triad dependence of the measured coupling without adjustable parameters.
This comparison reveals that the temperature-gradient and density-curvature contributions act at comparable strength and nearly cancel, jointly determining the coupling within the vortex.

\section*{Identification of a coherent monopolar vortex in PANTA}\label{monopole}
\par
\begin{figure}
    \centering
    \includegraphics[width=1\linewidth]{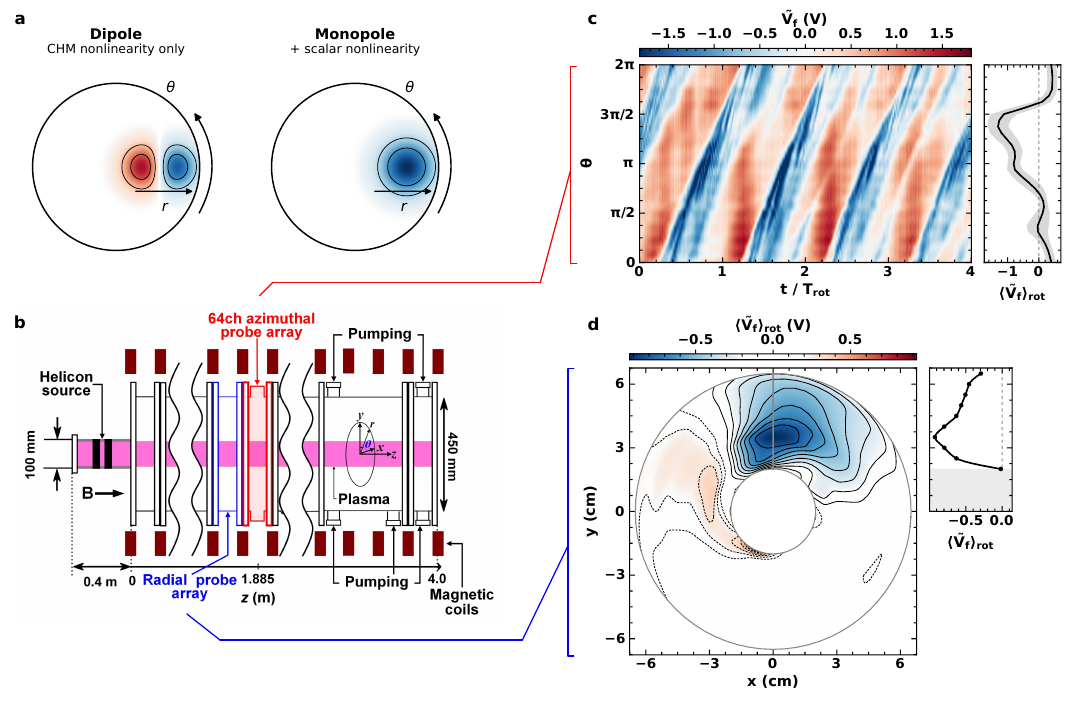}
    \caption{Dipolar versus monopolar vortices, and a monopole in PANTA.
    \textbf{a}, Schematic potential structures of a dipolar vortex (modon) admitted by the Charney--Hasegawa--Mima (CHM) equation and of a monopolar vortex, which requires an additional scalar nonlinearity.
    For azimuthal propagation, the oppositely signed lobes of the dipole are separated radially, perpendicular to the propagation direction.
    \textbf{b}, Schematic of PANTA and the diagnostic arrangement.
    Red and blue lines connect the azimuthal and radial Langmuir-probe arrays to the data in \textbf{c} and \textbf{d}, respectively.
    \textbf{c}, Azimuth--time evolution of the floating-potential fluctuation $\widetilde V_{\rm f}$ at $r=4$ cm and an axial magnetic field of 90 mT (discharge A).
    Time is normalized by the vortex rotation period $T_{\rm rot}=2\pi/\Omega_{\rm v}$, where $\Omega_{\rm v}$ is the angular velocity of the vortex.
    Right, co-rotating waveform $\langle\widetilde V_{\rm f}\rangle_{\rm rot}$ from the same discharge, where $\langle\cdot\rangle_{\rm rot}$ denotes the average over successive rotations at fixed co-rotating angle $\xi=\theta-\Omega_{\rm v}t$; the waveform is referenced to its median over $\xi$ and plotted against $\theta$ at $t=0$, and the grey band denotes one standard deviation across rotations.
    \textbf{d}, Two-dimensional structure of $\langle\widetilde V_{\rm f}\rangle_{\rm rot}$ reconstructed from the radial array in a separate discharge with identical operating parameters (discharge B), referenced at each radius to its median over $\xi$, with $\xi$ as the polar angle measured anticlockwise from the positive $y$ axis.
    The colour scale is white for $|\langle\widetilde V_{\rm f}\rangle_{\rm rot}|<0.2$ V; solid and dashed contours denote negative and positive values at intervals of 0.1 V, and the central region $r<2$ cm was not measured.
    Right, radial profile along $\xi=0$ (vertical line) at the radial-array channels; the grey band marks the unmeasured region.
    The colour scale differs from that in \textbf{c}, which shows the instantaneous fluctuation rather than the rotation average.}
    \label{fig:1}

\end{figure}

Before testing drift-wave theory against the coupling inside the vortex, we establish that the observed structure is a coherent monopole.
Its components must rotate together as a single pattern, and its potential must have one core rather than the two oppositely signed lobes of a dipole.
Experiments were performed in PANTA, where a helicon source produces an argon plasma column confined by a uniform axial magnetic field (Fig.~\ref{fig:1}b).
Floating-potential fluctuations were measured using 32 alternating probes in a 64-probe azimuthal Langmuir-probe array~\cite{oldenburger2012,yamada2007}.
We focus first on $B_{\mathrm{ax}}=90$ mT, where the coherent structure is most clearly resolved.
\par
The azimuth--time evolution reveals a single coherent pattern that repeatedly circulates around the plasma with an approximately constant angular velocity (Fig.~\ref{fig:1}c).
In the frequency--azimuthal-mode-number spectrum, the dominant coherent component of each azimuthal mode number $m$ lies at a frequency $f_m\simeq m f_{\mathrm v}$ (see Fig.~\ref{fig:2}b, 90 mT), where $f_{\mathrm v}=1.22$ kHz is the common rotation frequency of the pattern and $\Omega_{\mathrm v}=2\pi f_{\mathrm v}$ its angular velocity.
This common-velocity relation, rather than the mere presence of harmonics, indicates that the components are stationary in one co-rotating frame, and the recurrence of the pattern over several rotations shows that their relative phases remain stable.
Transforming these floating-potential signals into the co-rotating coordinate ($\xi=\theta-\Omega_{\mathrm v}t$) and averaging over successive rotations yields an asymmetric waveform dominated by a single negative excursion (Fig.~\ref{fig:1}c, right).

\par
We next examine whether this coherent waveform has monopolar or dipolar structure.
A single-radius waveform alone cannot exclude a dipolar vortex, whose oppositely signed lobes are displaced perpendicular to its propagation direction, that is, radially for an azimuthally propagating structure (Fig.~\ref{fig:1}a).
For this purpose, floating-potential fluctuations were measured with a radial probe array over $2\le r\le6.5$ cm in a separate discharge with identical operating parameters, in which the azimuthal array yielded the same $\Omega_{\mathrm v}$ and a co-rotating waveform of the same negative polarity.
Repeating the transformation at each radius reconstructs the co-rotating structure in two dimensions (Fig.~\ref{fig:1}d).
This two-dimensional structure consists of a single core with no oppositely signed structure at any radius along its azimuth, and the radial profile through the core does not change sign over the measured range (Fig.~\ref{fig:1}d, right).
Its weak positive regions, like the positive excursion of the waveform in Fig.~\ref{fig:1}c, are displaced azimuthally from the core rather than radially, and therefore do not constitute a radially displaced lobe.

\par
The common rotation velocity, persistent phase relation and single-core polarity structure together support the identification of the 90 mT state as a coherent monopolar vortex within the diagnostic resolution.
To identify what couples its harmonics, we next analyse this vortex with drift-wave theory.

\section*{An extended drift-wave equation}
\par
Existing reduced equations for monopolar drift vortices extend the Charney--Hasegawa--Mima equation by a single scalar nonlinearity, from either the temperature gradient or the density-profile curvature, depending on the asymptotic ordering adopted~\cite{laedkespatschek1986,laedke1988,lakhin1988,spatschek1990,salat1990,salat1991}.
They also treat the density scale length as long compared with the ion-sound gyroradius $\rho_s$.
In PANTA, the density scale length is comparable to $\rho_s$, and the electron-temperature gradient and the density-profile curvature are both present.
We therefore rederive the drift-wave equation with an ordering suited to these conditions.
In the resulting equation, both nonlinearities appear at the same order, with coefficients evaluated from the measured equilibrium profiles.
From the same profiles, we then examine whether the vortex propagates outside the linear drift-wave continuum, as a long-lived monopole requires.
\par
\begin{figure}
    \centering
    \includegraphics[width=1\linewidth]{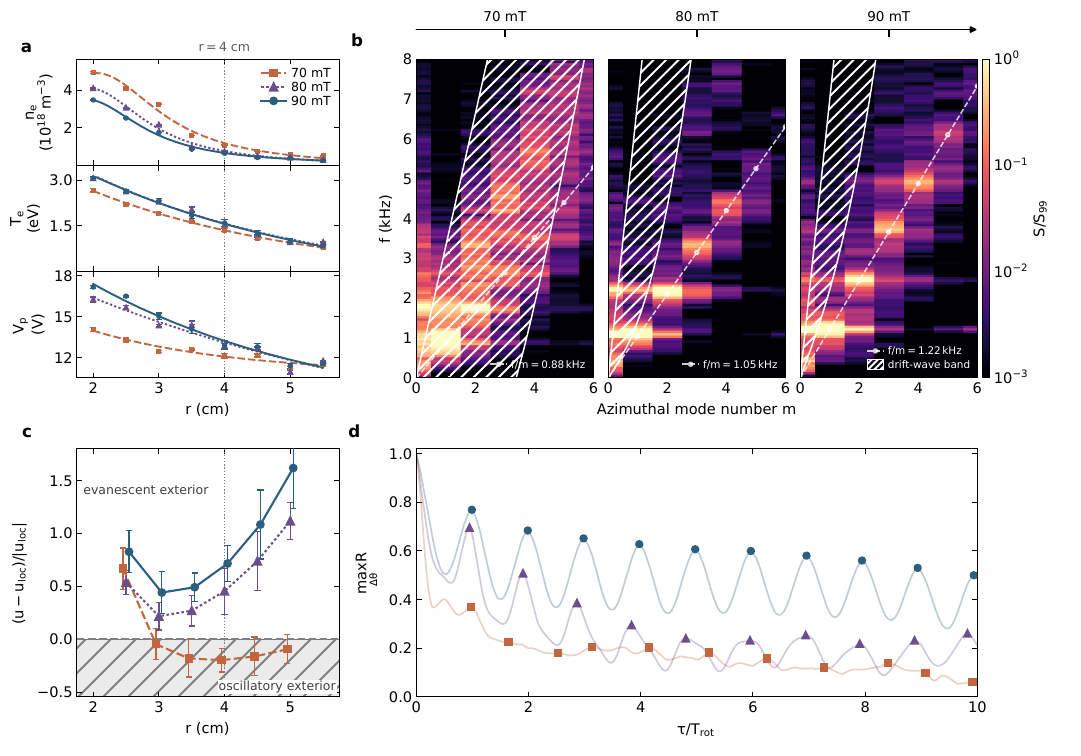}
    \caption{Equilibrium profiles and the drift-wave continuum at three field settings.
    \textbf{a}, Radial profiles of the electron density $n_e$, electron temperature $T_e$ and plasma potential $V_{\rm p}$ at 70, 80 and 90 mT, obtained from swept Langmuir-probe current--voltage characteristics.
    Symbols denote measured values with one-standard-error bars, and curves denote the Lorentzian fits from which the radial gradients are evaluated (Methods).
    The vertical dotted lines in \textbf{a} and \textbf{c} mark $r=4$ cm, at which the drift-wave continuum in \textbf{b} is evaluated.
    \textbf{b}, Frequency--azimuthal-mode-number spectra at 70, 80 and 90 mT (increasing along the arrow), each normalized by its 99th-percentile spectral density $S_{99}$.
    The drift-wave continuum is evaluated locally at $r=4$ cm using the equilibrium profiles in \textbf{a}.
    The measured $E\times B$ Doppler shift is included in the calculated laboratory-frame continuum.
    White circles, dashed lines and hatching denote spectral peaks, the common-velocity relation $f_m=m f_{\rm v}$ and the calculated drift-wave continuum, respectively.
    \textbf{c}, Radial localization margin $(u-u_{\rm loc})/|u_{\rm loc}|$, evaluated for the $m=1$ component, where $u=(r\Omega_{\rm v}-v_{E\times B})/c_s$ is the vortex velocity in the plasma frame.
    Both $u$ and $u_{\rm loc}$ are evaluated using the same velocity normalization, which cancels in the plotted ratio.
    Error bars show propagated one-standard-deviation uncertainties from the measured rotation velocity and equilibrium-profile parameters.
    Hatching marks $0<u<u_{\rm loc}$, where the exterior is oscillatory.
    \textbf{d}, Maximum over $\Delta\theta$ of the normalized azimuth--time correlation $R(\Delta\theta,\tau)$ (Methods) versus time lag normalized by $T_{\rm rot}$.
    Pale curves show the full correlation maxima, symbols identify successive recurrence peaks.
    Colours and symbols follow \textbf{a}.}
    \label{fig:2}
\end{figure}

Assuming cold ions, Boltzmann electrons and negligible parallel ion dynamics, we solve the ion momentum equation iteratively in the low-frequency parameter $\omega/\omega_{\mathrm{ci}}\ll1$, without expanding in the fluctuation amplitude, and combine the resulting velocity with ion continuity~\cite{hasegawa1978,nycander1991large}.
To leading order in $k_y\rho_s$ and in the radial decay rate of the potential outside the vortex core (Methods), the resulting potential-vorticity equation reads
\begin{equation}
\begin{aligned}
\left(1-\nabla^2-\kappa_n\partial_x\right)\partial_t\phi
=
\kappa_n\partial_y\phi
-\kappa_T\phi\partial_y\phi
+\kappa_n\left\{\phi,\partial_x\phi\right\}
+\left\{\phi,\nabla^2\phi\right\}.
\end{aligned}
\label{eq:drift}
\end{equation}
Here, $\phi$ denotes the normalized electrostatic potential fluctuation, whereas $x$ and $y$ denote the local radial and azimuthal coordinates, respectively.
We define the normalized equilibrium gradients as $\kappa_n=\partial_x\ln n_0$ for density and $\kappa_T=\partial_x\ln T_{e0}$ for electron temperature, and the two-dimensional Poisson bracket as $\{f,g\}\equiv(\partial_x f)(\partial_y g)-(\partial_y f)(\partial_x g)$.
Equilibrium profiles were measured at $B_{\mathrm{ax}}=70$, 80 and 90 mT (Fig.~\ref{fig:2}a), and $\kappa_n$ and $\kappa_T$ are evaluated from them for each field.
The right-hand side comprises the linear drift-wave term, the temperature-gradient scalar nonlinearity, and the diamagnetic and Hasegawa--Mima Jacobian nonlinearities.
\par
Besides a scalar nonlinearity, a monopolar solution of equation~\eqref{eq:drift} requires a propagation velocity outside the linear drift-wave continuum.
Inside the continuum, the vortex resonates with linear drift waves of the same phase velocity and loses energy by radiating them, so it cannot persist.
For a steadily translating solution $\phi(x,y-ut)$ with plasma-frame velocity $u$, the quadratic terms are negligible outside the nonlinear core~\cite{su1991,nycander1991large}, so the exterior obeys the linearized equation.
Its potential then decays with radius, rather than oscillating, only if
\begin{equation}
u<0\quad\text{or}\quad
u>u_{\mathrm{loc}}
\equiv
-\frac{\kappa_n}
{1+k_y^2+\kappa_n^2/4}.
\label{eq:localization}
\end{equation}
Here $u_{\mathrm{loc}}$ marks the edge of the linear drift-wave phase-velocity continuum.
We refer to equation~\eqref{eq:localization} as the localization condition, because it confines the vortex potential to the core instead of letting it radiate; it is necessary but not sufficient for a localized vortex (Methods).
\par
The two lower fields, which correspond to distinct equilibria, provide reference states for the 90 mT vortex.
At 90 mT, the dominant spectral peaks form a common-velocity harmonic branch separated from the drift-wave continuum calculated from the measured profiles (Fig.~\ref{fig:2}b), and the margin $(u-u_{\mathrm{loc}})/|u_{\mathrm{loc}}|$, with $u=\left(r\Omega_{\mathrm v}-v_{E\times B}\right)/c_s$, is positive at all measured radii (Fig.~\ref{fig:2}c).
At 80 mT the margin remains positive but is smaller, with the branch closer to the continuum boundary, and at 70 mT the low-mode peaks overlap the continuum and the margin is negative at the measurement radius $r=4$ cm.
The persistence of the rotating structure follows the same sequence: the recurrence peaks of the normalized azimuth--time correlation $R(\Delta\theta,\tau)$ decay within a rotation at 70 mT, more gradually at 80 mT and remain high over the measured interval at 90 mT (Fig.~\ref{fig:2}d), with the correlation ridge propagating at the measured $\Omega_{\mathrm v}$ (Extended Data Fig.~\ref{fig:excorr}).
In the same sequence, the three-wave interactions among the harmonics become less closely frequency-matched and less phase-coherent from 90 to 70 mT (Extended Data Fig.~\ref{fig:ex1}), as expected when the harmonics are no longer locked to a single rotating structure.
Only the 90 mT state thus combines a velocity outside the continuum, long-lived rotation and phase-locked harmonics, and we use it to test the coupling predicted by equation~\eqref{eq:drift}.

\section*{A tug-of-war between two nonlinearities}
\par
Equation~\eqref{eq:drift} predicts the strength of each three-wave interaction among the vortex harmonics, in which two harmonics $m_1$ and $m_2$ combine to drive a third, $m_3=m_1+m_2$ (a triad).
Each candidate nonlinearity enters this prediction as a separate term, so comparing it with the measured coupling both tests the equation and resolves the individual contributions.
\par
The simultaneous probe measurements allow the floating-potential fluctuations to be decomposed directly into azimuthal Fourier components $X_m(t)$, without invoking a frequency--wavenumber mapping.
For each target mode $m_3$, we represent its evolution as a linear response plus quadratic interactions over triads $j\equiv(m_3;m_1,m_2)$ satisfying $m_3=m_1+m_2$, and infer the corresponding coefficients from the simultaneous mode time series.
Defining the measured bispectrum as $B_j=\langle X_{m_3}^{*}X_{m_1}X_{m_2}\rangle$~\cite{kim1979,yamada2010}, the mode evolution and the nonlinear contribution to modal fluctuation power are
\begin{equation}
\begin{aligned}
\frac{\partial X_{m_3}}{\partial t}
&=
\Lambda_{m_3}^{L}X_{m_3}
+
\sum_{m_1+m_2=m_3}
\Lambda_j^{Q}X_{m_1}X_{m_2},\\
\left.\frac{\partial P_{m_3}}{\partial t}\right|_{\rm NL}
&=
\sum_{m_1+m_2=m_3}T_j,
\qquad
T_j\equiv2\operatorname{Re}\left(\Lambda_j^{Q}B_j\right).
\end{aligned}
\label{eq:nonlinear_transfer}
\end{equation}
Here, $P_{m_3}=\langle|X_{m_3}|^2\rangle$ is the mean modal fluctuation power, $\Lambda_{m_3}^L$ and $\Lambda_j^{Q}$ are the linear and quadratic coefficients, and $T_j$ is the signed nonlinear power transfer rate into the target mode $m_3$ through triad $j$.
The coefficients are estimated using an established quadratic system-identification method~\cite{ritz1986,ritz1989,kimjs1996,dewit1999}.
Details of our Schur-complement implementation and the resolution criterion used to retain 29 experimentally resolvable triads are provided in Methods.
\par
\begin{figure}
    \centering
    \includegraphics[width=1\linewidth]{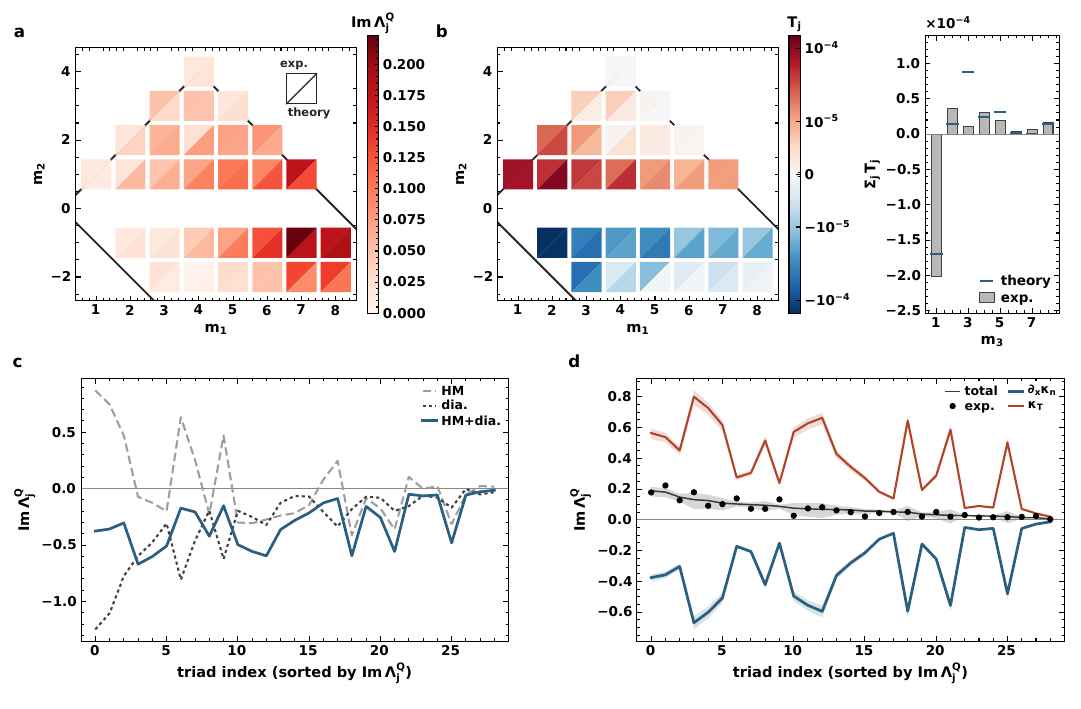}
    \caption{Measured three-wave coupling compared with the extended drift-wave equation.
    \textbf{a}, Triad-resolved quadratic coupling coefficients $\Lambda_j^Q$ at 90 mT, with $j\equiv(m_3;m_1,m_2)$.
    Upper-left and lower-right triangles show the experimentally inferred and theoretical coefficients, respectively, on a common colour scale.
    The theoretical coefficients were forward-projected through the resolution matrix described in Methods.
    \textbf{b}, Corresponding signed nonlinear power transfer $T_j$ at 90 mT.
    The split-tile convention follows \textbf{a}.
    The accompanying graph shows the transfer summed over retained triads for each target mode $m_3$; grey bars and blue ticks denote the experimental and theoretical values, respectively.
    \textbf{c}, Hasegawa--Mima polarization (HM) and density-gradient diamagnetic (dia.) contributions to the forward-projected theoretical $\operatorname{Im}\Lambda_j^Q$ at 90 mT and their sum (HM+dia.), across the retained triads sorted by the total theoretical $\operatorname{Im}\Lambda_j^Q$.
    \textbf{d}, Temperature-gradient ($\kappa_T$) and density-curvature ($\partial_x\kappa_n$) contributions and their total, with the experimentally inferred coefficients (filled circles), for the same triads and ordering as \textbf{c}.
    Shaded bands span the 16th to 84th percentiles of each theoretical curve under resampling of the equilibrium-profile measurements (Methods).
    The $\partial_x\kappa_n$ curve is the HM+dia. sum in \textbf{c}, which reduces approximately to the density-curvature form of equation~\eqref{eq:jacobian}.}
    \label{fig:3}
\end{figure}

The experimentally inferred coupling coefficients $\Lambda_j^Q$ agree with those predicted by equation~\eqref{eq:drift} in sign and triad-to-triad variation across the retained triads (Fig.~\ref{fig:3}a), with a correlation coefficient of 0.91 between the imaginary parts, and in magnitude without adjustment (Methods).
Through these couplings, the $m_3=1$ component loses power and the higher harmonics gain it (Fig.~\ref{fig:3}b), in both the measurement and the prediction.
The same identification shows that the fundamental is linearly driven and the harmonics are linearly damped (Methods).
Without this transfer, the harmonics would therefore decay, and the waveform would relax towards a sinusoidal $m=1$ wave with equal positive and negative excursions.
The nonlinear coupling thus builds the single-core waveform by transferring energy from the fundamental into the harmonics, rather than merely locking the phases of independently driven harmonics.

\par
To resolve the respective roles of these nonlinearities, we decompose the predicted coefficient into its temperature-gradient and Jacobian contributions (Fig.~\ref{fig:3}d), the latter being the sum of the diamagnetic and Hasegawa--Mima terms (Fig.~\ref{fig:3}c).
For radial mode functions satisfying the linearized traveling-wave equation with a common propagation velocity, the combined Jacobian contribution can be expressed as 
\begin{equation}
    \kappa_n\{\phi,\partial_x\phi\}+\{\phi,\nabla^2\phi\}\simeq(\partial_x\kappa_n)\left(\partial_x\phi -\phi /u\right)\partial_y\phi.
\label{eq:jacobian}
\end{equation}
This reduction applies to a localized common-velocity state, which Fig.~\ref{fig:2} identifies clearly only at 90 mT.
It also involves the radial structure of the mode functions, whereas the temperature-gradient term contains no radial derivative of $\phi$.
The Hasegawa--Mima and diamagnetic contributions are individually large and vary strongly from triad to triad (Fig.~\ref{fig:3}c), but their sum is much smaller (Fig.~\ref{fig:3}d), as equation~\eqref{eq:jacobian} anticipates.
This sum closely follows the triad dependence of the temperature-gradient contribution with the opposite sign (Fig.~\ref{fig:3}d), which indicates that the scalar component proportional to $\phi\partial_y\phi$ dominates over $(\partial_x\phi)(\partial_y\phi)$.
The combined Jacobian nonlinearity can therefore be treated as an effective scalar nonlinearity driven by density-profile curvature, and it is this combination, rather than either Jacobian term separately, that the measurements can constrain.
It is the density-curvature nonlinearity identified in earlier reductions of the drift-wave equations~\cite{lakhin1987,lakhin1988}, which thus appears here alongside the temperature-gradient term.
\par
The temperature-gradient and density-curvature contributions are comparable in magnitude and oppositely signed throughout the retained triads (Fig.~\ref{fig:3}d).
Either contribution alone would exceed the measured coupling several-fold or reverse its sign, so the measured coupling is the small residue of their largely cancelling contributions.
Because the degree of cancellation varies from 63\% to 96\% across the triads, this residue has a triad pattern that neither contribution reproduces alone, whereas their sum does (Methods).
The temperature gradient sets the positive sign of the coupling, and the competing curvature contribution, through the triad-dependent degree of cancellation, shapes its magnitude across the harmonics.
Both contributions have appeared in earlier theories, which either retained them together under a special cancellation condition~\cite{su1991,nycander1989temp,nycander1989exist} or selected one of them through the choice of asymptotic ordering~\cite{lakhin1988,spatschek1990,salat1990,salat1991}.
In the measured vortex, neither can be neglected.
\par
These measurements provide the first direct comparison between drift-wave theory and nonlinear coupling measured within a monopolar drift vortex, showing that the extended drift-wave equation reproduces the observed coupling without adjustable parameters.
The measured nonlinear transfer supplies fluctuation power from the linearly driven fundamental to damped harmonics, sustaining the single-core waveform.
Within the extended equation, both the electron-temperature gradient and density-profile curvature are essential to this process.
Their opposing contributions leave a net coupling much smaller than either contribution alone, making the transfer that maintains the waveform sensitive to changes in either equilibrium profile.
This sensitivity motivates investigating whether profile variations can modify vortex-mediated transport, particularly at the edge of fusion plasmas, where steep temperature gradients and curved density profiles coexist.
More broadly, comparing measured interactions with the individual nonlinear terms of a governing equation offers a way to distinguish mechanisms that produce similar coherent structures.
Extending this approach to rotating-fluid experiments and simulations of planetary flows could reveal how competing nonlinearities maintain vortices across different physical systems.

\bibliography{monopolar}
\newpage
\section*{Methods}\label{sec:methods}
\subsection*{Plasma production, diagnostics and equilibrium}
\par
Experiments were performed in PANTA~\cite{oldenburger2012}.
The helicon source was operated at 7 MHz with an input power of 3 kW. The argon pressure, measured in neutral gas without the magnetic field, was 0.466 Pa, and the gas flow rate was 50 sccm.
The axial coordinate $z$ is measured downstream from the helicon source.
Experiments were performed at $B_{\rm ax}=70$, 80 and 90 mT, with the other discharge parameters held fixed.
For each setting, realizations were started within a 250 ms interval of the stationary phase (0.300--0.550 s).
\par
Floating-potential fluctuations were measured at $r=4$ cm and $z=1.885$ m using 32 alternating probes in a 64-probe azimuthal Langmuir-probe array~\cite{oldenburger2012}.
The signals were recorded simultaneously at 1 MHz.
A radially movable Langmuir probe array located at $z=1.635$ m was used to measure the equilibrium profiles from swept current--voltage characteristics.
The channels analysed here spanned $r=2$--5.5 cm with a spacing of 0.5 cm.
The azimuthal probe array precludes profile measurements at the same axial position, and the equilibrium profiles measured 0.25 m upstream at $z=1.635$ m were therefore used as proxies for those at the fluctuation-measurement plane ($z=1.885$ m).
The same radial array was also used to measure the radial fluctuation structure shown in Fig.~\ref{fig:1}d, as described below.

\par
At $B_{\rm ax}=90$ mT, two discharges with identical operating parameters were analysed.
Discharge A provided the data for Fig.~\ref{fig:1}c and Figs.~\ref{fig:2} and \ref{fig:3}.
In discharge B, the 10-channel radial array was operated in floating-potential mode over $r=2$--6.5 cm with a spacing of 0.5 cm and recorded at 1 MHz, so that the radial fluctuation structure in Fig.~\ref{fig:1}d was measured at the same axial position as the equilibrium profiles.
The azimuthal probe array, operated simultaneously, yielded the same rotation frequency as in discharge A, $f_{\rm v}=1.22$ kHz.

\par
For each magnetic field setting, the bias voltage applied to the radial Langmuir probe was swept from $-50$ to $50$ V at 50 Hz.
The electron temperature was obtained from the electron-current branch, the plasma potential was identified from the zero crossing of $\mathrm{d}^2I/\mathrm{d}V^2$, and the electron density was inferred from the ion-saturation current under the Bohm-sheath approximation.
The radial electric field and azimuthal $\mathbf{E}\times\mathbf{B}$ velocity were calculated as
\begin{equation}
E_r=-\frac{\partial V_{\rm p}}{\partial r},
\qquad
v_{E\times B}=-\frac{E_r}{B_{\rm ax}}.
\end{equation}
Before differentiation, the measured $n_0(r)$, $T_{e0}(r)$ and
$V_{\rm p}(r)$ profiles (Fig.~\ref{fig:2}a) were fitted with Lorentzian functions. Radial derivatives were evaluated from the fitted profiles, which served as equilibrium inputs for the fluctuation analysis at $z=1.885$ m and were used to evaluate $\rho_s$, $\kappa_n$, $\kappa_T$ and the linear drift-wave continuum without fitting to the fluctuation data.
Profile uncertainties are reported as one standard error of the mean.

\subsection*{Mode-resolved analysis and vortex characterization}
\par
For each discharge, the temporal mean was subtracted from the floating-potential signals to obtain $\widetilde V_{\rm f}(\theta_j,t)$, where $\theta_j$ denotes the position of the $j$th probe.
The azimuthal-mode amplitudes were calculated from the 32 simultaneously measured signals as
\begin{equation}
X_m(t)=\frac{1}{N_\theta}
\sum_{j=0}^{N_\theta-1}
\widetilde V_{\rm f}(\theta_j,t)
\exp(-im\theta_j),
\qquad N_\theta=32.
\end{equation}
The frequency--azimuthal mode number spectrum $S(f,m)$ was obtained by applying a temporal Fourier transform to $X_m(t)$.
A 50 ms sub-interval of the stationary period was divided into five Hann-windowed segments of $2^{14}$ samples each (16.4 ms) with 50\% overlap.
This yielded a frequency-bin spacing of $\Delta=61$ Hz, and the resulting segment periodograms were averaged.
The dominant frequency $f_m$ of each resolved mode was identified from its spectral maximum, and the common rotation frequency $f_{\rm v}$ was obtained from a least-squares fit to $f_m=m f_{\rm v}$, constrained to pass through the origin.
The angular velocity and rotation period were then defined as $\Omega_{\rm v}=2\pi f_{\rm v}$ and $T_{\rm rot}=1/f_{\rm v}$, respectively.
For visualization, each spectrum was normalized by its 99th-percentile spectral density $S_{99}$.

\par
To characterize the vortex morphology, the floating-potential fluctuations measured with the azimuthal array in each discharge were reconstructed in the co-rotating coordinate $\xi=\theta-\Omega_{\rm v}t$ modulo $2\pi$ after subtracting the instantaneous azimuthal average $X_0(t)$.
Samples from successive rotations were grouped into 32 equally spaced phase bins.
The mean in each bin defines the co-rotating waveform, and its envelope denotes one standard deviation across the rotation ensemble.
The origin of $\xi$ was chosen such that the minimum of this waveform lies at $\xi=0$.
The polarity asymmetry of the fluctuations was quantified by the skewness, defined as the standardized third central moment, evaluated from the full time series of each probe after subtracting $X_0(t)$ and averaged over the 32 probes. The resulting skewness, $-0.69$ at 90 mT, involves no reference level and therefore confirms the negative polarity of the core independently of the reference described below.
The radial-array signals in discharge B, from which the temporal mean was subtracted, were transformed using the azimuthal position of the array, $\theta_{\rm p}$, and the same $\Omega_{\rm v}$, and binned in the same way at each radius, so that $\xi=0$ in Fig.~\ref{fig:1}d corresponds to the minimum of the azimuthal-array waveform recorded simultaneously in discharge B.

\par
Subtracting the temporal mean at a fixed position removes not only the equilibrium but also the azimuthal average of the vortex itself, so the axisymmetric component of the co-rotating structure is not determined by fluctuation measurements.
A zero azimuthal mean at each radius is therefore a convention, and it forces any single-signed core to be accompanied by an oppositely signed background at the same radius.
Since adding a function of $r$ alone leaves the azimuthal structure unchanged, we reference the waveform at each radius to its median over $\xi$, which approximates the ambient level away from the core without introducing an adjustable parameter.
The two-dimensional structure in Fig.~\ref{fig:1}d was obtained by interpolating the median-referenced waveforms between radial channels using piecewise cubic Hermite interpolation and is displayed with $\xi$ as the polar angle; the radial profile along $\xi=0$ shows the channel values without interpolation.
\par
The persistence of the rotating structure was quantified using the variance-normalized spatiotemporal autocorrelation
\begin{equation}
R(\Delta\theta,\tau_\ell)=
\frac{
\displaystyle\frac{1}{L-\ell}\sum_{j=0}^{N_{\rm\theta}-1}\sum_{n=0}^{L-\ell-1}
\widetilde V_j[n]\widetilde V_{(j+\Delta)\bmod N_{\rm\theta}}[n+\ell]
}{
\displaystyle\frac{1}{L}\sum_{j=0}^{N_{\rm\theta}-1}\sum_{n=0}^{L-1}\left[\widetilde V_j[n]\right]^2
},
\label{eq:spatiotemporal_correlation}
\end{equation}
where
\begin{equation}
\tau_\ell=\frac{\ell}{f_{\rm s}},
\qquad
\Delta\theta=\frac{2\pi\Delta}{N_{\rm\theta}}.
\end{equation}
Here $j$ indexes the $N_{\rm \theta}=32$ azimuthal channels, $L$ is the number of samples in the analysed time series, and $\ell$ and $\Delta$ are the temporal and azimuthal integer lags, respectively.
Channel indices were treated periodically. The correlation was evaluated over $0\leq\tau_{\ell}\leq10T_{\rm rot}$. At each time lag, the maximum azimuthal correlation was defined as
\begin{equation}
    R_{\max}(\tau_\ell)=
    \max_{\Delta\theta}R(\Delta\theta, \tau_\ell).
\end{equation}
Successive local maxima of $R_{\max}$ near integer multiples of $T_{\rm rot}$ were identified as recurrence peaks (Fig.~\ref{fig:2}d); the full correlation maps are shown in Extended Data Fig.~\ref{fig:excorr}.

\subsection*{Ordering and derivation of the extended drift-wave equation}
\par
We derive equation~\eqref{eq:drift} from the cold-ion fluid equations in a uniform magnetic field $\mathbf B=B\hat{\mathbf z}$, assuming quasi-neutrality, negligible parallel ion motion and Boltzmann electrons, $n=n_0(x)\exp[e\phi/T_e(x)]$, whose temperature varies across the field.
The ion momentum equation,
\begin{equation}
\partial_t\mathbf v+\mathbf v\cdot\nabla\mathbf v=-\frac{e}{m_i}\nabla\phi+\omega_{\rm ci}\,\mathbf v\times\hat{\mathbf z},
\end{equation}
is solved iteratively in the small parameter $\epsilon_\omega\sim\omega_{\rm ci}^{-1}\partial_t\sim\omega_{\rm ci}^{-1}\mathbf v\cdot\nabla\ll1$.
At lowest order the ion velocity is the $\mathbf E\times\mathbf B$ drift $\mathbf v_0=(e/m_i\omega_{\rm ci})\,\hat{\mathbf z}\times\nabla\phi$, with vorticity $\Omega_0=(e/m_i\omega_{\rm ci})\nabla^2\phi$, and the next-order correction is
\begin{equation}
\mathbf v_1=
-\frac{e}{m_i\omega_{\rm ci}^2}\nabla\partial_t\phi
+\frac{e^2}{m_i^2\omega_{\rm ci}^2}\,\hat{\mathbf z}\times\nabla\frac{|\nabla\phi|^2}{2}
-\frac{\Omega_0}{\omega_{\rm ci}}\mathbf v_0,
\end{equation}
consisting of the linear and nonlinear polarization drifts and a vorticity correction, with $\Omega_1=\hat{\mathbf z}\cdot\nabla\times\mathbf v_1$.
Combining the curl of the momentum equation with ion continuity gives the potential-vorticity conservation law
\begin{equation}
\frac{\mathrm d}{\mathrm dt}\ln\frac{\omega_{\rm ci}+\Omega}{n}=0,
\end{equation}
into which $\mathbf v=\mathbf v_0+\mathbf v_1$, $\Omega=\Omega_0+\Omega_1$ and the Boltzmann density are substituted.
Time, perpendicular length, velocity and potential are normalized by $\omega_{\rm ci}^{-1}$, $\rho_s=c_s/\omega_{\rm ci}$, $c_s$ and $T_{e,\mathrm{ref}}/e$, respectively, where $c_s=(T_{e,\mathrm{ref}}/m_i)^{1/2}$.
About a local reference radius $r_0$, we define $x=(r-r_0)/\rho_s$, $y=r_0\theta/\rho_s$, $\kappa_n=\rho_s\partial_r\ln n_0$ and $\kappa_T=\rho_s\partial_r\ln T_{e0}$.
\par
After higher-order products such as $\mathbf{v}_1\cdot\nabla\Omega_1$ are neglected, the equation is ordered in two small parameters, with $\kappa_n$ and $u$ of order unity, the strength of the profile nonlinearities $\epsilon_{\rm p}\sim|\kappa_T|\sim\partial_x\kappa_n/u$ held fixed and no assumption on the fluctuation amplitude.
The first is the normalized azimuthal wavenumber $k_y$, which the drift expansion requires to be small because the ion-frame frequency is $\hat\omega/\omega_{\rm ci}=uk_y$.
The second is
\begin{equation}
D\equiv1+k_y^2+\frac{\kappa_n}{u},
\end{equation}
which sets the outward decay rate of the potential outside the core, $\phi'/\phi=|\kappa_n|/2-(D+\kappa_n^2/4)^{1/2}$ for $\kappa_n<0$.
In the limit $k_y,D\to0$ at fixed $\epsilon_{\rm p}$, the temperature-gradient and density-curvature nonlinearities are the leading nonlinear terms, the diamagnetic and Hasegawa--Mima nonlinearities combining into the curvature form of equation~\eqref{eq:jacobian}, whereas the advection of the fluctuating density and vorticity by the polarization drift enters at higher order and is neglected.
For the higher harmonics $k_y$ and $D$ are of order unity, and the comparison over the 29 retained triads therefore tests equation~\eqref{eq:drift} beyond its formal asymptotic range.
\par
In equation~\eqref{eq:drift}, the term $-\kappa_T\phi\partial_y\phi$ arises from $\mathbf E\times\mathbf B$ advection of the Boltzmann response $e\phi/T_e(x)$ across the temperature gradient, and $\{\phi,\nabla^2\phi\}$ is the advection of the vorticity, the Hasegawa--Mima nonlinearity.
The nonlinear polarization drift and the vorticity correction, acting on the density gradient, give $-\{|\nabla\phi|^2/2,\,-\ln n_0\}+\nabla^2\phi\,\{\phi,\,-\ln n_0\}=-\kappa_n\{\phi,\partial_x\phi\}$, the diamagnetic nonlinearity.
\par
Outside the nonlinear core, linearizing equation~\eqref{eq:drift} in the traveling-wave coordinate $\eta=y-ut$ with locally constant coefficients gives the radial wavenumbers
\begin{equation}
k_x^\pm=
\frac{i\kappa_n}{2}
\pm i
\left(
1+k_y^2+\frac{\kappa_n^2}{4}+\frac{\kappa_n}{u}
\right)^{1/2}.
\end{equation}
For $\kappa_n<0$, the radial response is non-oscillatory only if $u<0$ or $u>u_{\rm loc}$ (equation~\eqref{eq:localization}); otherwise it belongs to the linear drift-wave continuum. This condition excludes radiation into the continuum, but it is not sufficient for a localized vortex because it does not establish that a nonlinear core solution exists or that the exterior decays over the measured radial range.
For Fig.~\ref{fig:2}b,c, $k_y=m\rho_s/r$ and $u=(r\Omega_{\rm v}-v_{E\times B})/c_s$ were evaluated from the measured profiles, the continuum was obtained by varying $q_x=\operatorname{Re}k_x$ and applying the measured $\mathbf{E}\times\mathbf{B}$ Doppler shift, and the margin was evaluated for $m=1$, the most restrictive harmonic, with one-standard-deviation uncertainties propagated from the rotation velocity and profile parameters.

\subsection*{Quadratic system identification and nonlinear transfer}
\par
The azimuthal-mode amplitudes $X_m(t)$ defined above were used to estimate the quadratic coupling coefficients from the simultaneous probe measurements.
For this analysis, $e\widetilde V_{\rm f}/T_{e0}$ was used as an experimental proxy for $\phi$; its sensitivity to electron-temperature fluctuations is examined at the end of this section.
Each stationary record was divided into 50\%-overlapping realizations of duration $16T_{\rm rot}$, starting within the 250 ms interval defined above, so that the analysed records extend over 262--267 ms depending on the field setting.
Within each realization, a slowly varying background was subtracted from each channel using a zero-phase Gaussian smoother with a standard deviation of $T_{\rm rot}/\sqrt{3}$, corresponding to a box-equivalent width of $2T_{\rm rot}$.
The smoothing width and realization duration were determined separately for each magnetic-field setting, since $T_{\rm rot}$ depends on $B_{\rm ax}$. Mode amplitudes separated by a time lag $\tau=10~\mu{\rm s}$ were related through the discrete quadratic model
\begin{equation}
Y=LX+\sum_{j\in\mathcal J_{m_3}}Q_j z_j+\varepsilon,
\qquad
X=X_{m_3}(t),\quad
Y=X_{m_3}(t+\tau),\quad
z_j=X_{m_1}(t)X_{m_2}(t),
\label{eq:discrete_quadratic_model}
\end{equation}
where $\mathcal J_{m_3}$ contains the triads $j\equiv(m_3;m_1,m_2)$ satisfying $m_1+m_2=m_3$, and $\varepsilon$ denotes the part of the delayed signal not represented by the linear and quadratic predictors~\cite{ritz1986,ritz1989,kimjs1996,dewit1999}. Parent pairs were ordered as $m_1\geq m_2$ to avoid counting the same quadratic product twice.
\par
The linear and quadratic coefficients were obtained from the corresponding normal equations.
Defining the vector of quadratic predictors $\mathbf z=(z_1,\ldots,z_J)^{\mathsf T}$ and
\begin{equation}
\begin{gathered}
P=\langle |X|^2\rangle,\qquad
d=\langle X^*Y\rangle,\qquad
\mathbf u=\langle\mathbf z^*X\rangle,\\
\mathbf A=\langle\mathbf z^*\mathbf z^{\mathsf T}\rangle,\qquad
\mathbf c=\langle\mathbf z^*Y\rangle,
\end{gathered}
\end{equation}
where the brackets denote a temporal average within each realization, the quadratic coefficient vector $\mathbf q=(Q_1,\ldots,Q_J)^{\mathsf T}$ was calculated using the Schur complement,
\begin{equation}
\begin{aligned}
\mathbf S&=\mathbf A-\frac{\mathbf u\mathbf u^\dagger}{P},\\
\mathbf q&=(\mathbf S+\lambda\mathbf I)^{-1}
\left(\mathbf c-\frac{\mathbf u d}{P}\right),\\
L&=\frac{d-\mathbf u^\dagger\mathbf q}{P}.
\end{aligned}
\label{eq:schur_identification}
\end{equation}
The diagonal regularization parameter $\lambda$ was determined separately for each target mode using a discrepancy criterion based on a Bartlett-kernel estimate of the finite-record sampling error. This regularization stabilizes the inversion when the quadratic predictors are strongly correlated.
\par
With $\tau$ much shorter than the characteristic evolution time, the discrete coefficients were converted to first-order continuous-time coefficients by removing the phase advance contained in the linear propagator:
\begin{equation}
\Lambda_{m_3}^{L}
=
\frac{|L|-1+i\arg(L)}{\tau},
\qquad
\Lambda_j^{Q}
=
\frac{|L|}{L}\frac{Q_j}{\tau}.
\label{eq:continuous_coefficients}
\end{equation}
The factor $|L|/L=\exp[-i\arg(L)]$ removes the phase accumulated over the finite time lag without changing the magnitude of the quadratic coefficient.
Candidate triads were constructed for target modes $1\leq m_3\leq8$, with $m_1\leq8$, $m_2\neq0$ and $m_1\geq m_2$.
Synthetic injection--recovery tests were used to determine the resolvable triad set before comparison with theory.
Triads with $m_2\leq-3$, for which less than $2\%$ of the injected coefficient amplitude was recovered, were excluded as poorly resolved.
Restricting the analysis to $m_2\geq-2$ yielded 29 retained triads.
The injection--recovery analysis identified the resolvable triad set and validated the resolution matrix used in the theoretical comparison.
\par
For each retained triad, the bispectrum and nonlinear transfer were evaluated within the same realization from the discrete model of equation~\eqref{eq:discrete_quadratic_model}, following Ref.~\cite{kimjs1996},
\begin{equation}
B_j=\left\langle X_{m_3}^{*}X_{m_1}X_{m_2} \right\rangle,\qquad
T_j=\frac{1}{\tau}\operatorname{Re}\Bigl(2L^{*}Q_jB_j+Q_j\sum_{j'\in\mathcal J_{m_3}}Q_{j'}^{*}\left\langle z_jz_{j'}^{*}\right\rangle\Bigr).
\end{equation}
As $\tau\to0$, this finite-lag expression reduces to $T_j=2\operatorname{Re}(\Lambda_j^{Q}B_j)$ of equation~\eqref{eq:nonlinear_transfer}.
Positive and negative $T_j$ denote nonlinear power transfer into and out of the target mode, respectively.
The net nonlinear transfer was obtained by summing $T_j$ over the retained parent pairs.
The linear contribution to the modal power was evaluated in the same finite-lag form, $\tau^{-1}(|L|^2-1)P_{m_3}$, which reduces to $2\operatorname{Re}(\Lambda_{m_3}^{L})P_{m_3}$ as $\tau\to0$.
At 90 mT it is positive for $m_3=1$ and negative for $2\le m_3\le8$, so the fundamental is linearly driven, and the harmonics are linearly damped.
Both $\Lambda_j^Q$ and $T_j$ were evaluated separately in each realization and then averaged over all realizations, with none excluded.
Coupling coefficients and transfer rates were normalized by $\omega_{\rm ci}$.

\par
Because the floating potential also responds to electron-temperature fluctuations, $e\widetilde V_{\rm f}/T_{e0}=\tilde\phi-\alpha\widetilde T_e/T_{e0}$ with $\alpha=\ln[(m_i/2\pi m_e)^{1/2}]\approx4.7$ for argon, and $\widetilde T_e$ was not measured, the identification at 90 mT was repeated with the 32 ion-saturation-current probes interleaved in the same array.
Normalized by its temporal mean, this signal gives $\tilde I_{\rm sat}/I_{\rm sat0}=\tilde n/n_0+\widetilde T_e/2T_{e0}$, which, for the Boltzmann electrons assumed in equation~\eqref{eq:drift}, is a second proxy for $\tilde\phi$ with a temperature sensitivity smaller by $2\alpha\approx9$.
With identical analysis settings and resolution matrices computed separately for each signal, the coefficients obtained from the two signals were correlated with $r=0.75$ across the 29 retained triads, and both reproduced the triad dependence of the forward-projected theory described in `Theoretical coupling and comparison with experiment' ($r=0.91$ for $V_{\rm f}$ and $0.88$ for $I_{\rm sat}$; Extended Data Fig.~\ref{fig:exisat}).
The agreement with theory is therefore not specific to the floating-potential proxy, and the triad structure of the $V_{\rm f}$-based coefficients is not dominated by the electron-temperature contribution.

\subsection*{Theoretical coupling and comparison with experiment}
\par
The theoretical coefficients $\Lambda_j^{Q,\mathrm{th}}$ were calculated from equation~\eqref{eq:drift} using an azimuthal Fourier representation with numerically determined radial mode functions.
Every nonlinear term in equation~\eqref{eq:drift} carries a single azimuthal derivative, which makes the predicted coefficients purely imaginary; the real parts of the measured coefficients, which represent amplitude growth or decay within the triad, are not predicted by the equation and are not compared.
For each azimuthal mode, the radial structure was obtained by integrating the linearized traveling-wave equation $\phi_m''+\kappa_n(x)\phi_m'-D_m(x)\phi_m=0$ inward from $x=6$, where $\phi_m'/\phi_m$ was set to the outward-decaying root of the local constant-coefficient equation, and primes denote derivatives with respect to the normalized radial coordinate.
Here $x=0$ corresponds to the measurement radius $r=4$ cm, $\kappa_n(x)=\kappa_{n0}+\kappa_{n0}'x$, and $D_m(x)=1+k_{y,m}^2+\kappa_n(x)/u$.
The coefficients $\kappa_{n0}$ and $\kappa_{n0}'$ were evaluated from the measured equilibrium-profile fits, while $k_{y,m}$ and $u$ were held at their values at the measurement radius.
The radial mode functions were normalized at $x=0$, and their derivatives at that location were used to evaluate the quadratic coupling coefficients by collecting products satisfying $m_1+m_2=m_3$.
This calculation retains the radial variation of the density gradient in the mode structure and its contribution to the Jacobian nonlinearities.
The temperature-gradient, diamagnetic and Hasegawa--Mima contributions were evaluated separately and then summed.
The uncertainty of the theoretical coefficients due to the equilibrium profiles was evaluated by an exact bootstrap over the five radial sweeps whose average defines the profiles.
For each of the 126 distinct resamples, weighted by its multinomial probability, the profiles were refitted, $T_e$, $\kappa_n$, $\kappa_T$, $u$ and $\partial_x\kappa_n$ at $r=4$ cm were re-evaluated, and the coefficients were recomputed and forward-projected.
The change of each parameter relative to the fit to all five sweeps was added to its nominal value, so that the uncertainty bands, spanning the probability-weighted 16th to 84th percentiles, are centred on the nominal prediction.
\par
For each target mode, the resolution matrix was calculated from the corresponding Schur matrix and regularization parameter as
\begin{equation}
    \mathbf R_{m_3}=(\mathbf{S_{m_3}}+\lambda_{m_3}\mathbf{I})^{-1}\mathbf S_{m_3}.
\end{equation}
Before comparison with experiment, the theoretical coefficient vector for each target mode was forward-projected as
\begin{equation}
    \boldsymbol{\Lambda}_{m_3}^{Q,\mathrm{th,obs}}=
    \mathbf R_{m_3}\boldsymbol{\Lambda}_{m_3}^{Q,\mathrm{th}}.
\end{equation}
The same projection was applied separately to each nonlinear contribution.
The theoretical coefficients are continuous-time quantities, and the theoretical transfer was therefore calculated directly from this limiting form using the experimentally measured bispectrum,
\begin{equation}
    T_j^{\mathrm{th}}=
    2\operatorname{Re}\left(
    \Lambda_j^{Q,\mathrm{th,obs}}B_j^{\mathrm{exp}}
    \right).
\end{equation}
Agreement between the imaginary parts of the experimental and theoretical coefficients was quantified by their correlation coefficient over the 29 retained triads.
\par
The prediction also reproduces the measured magnitudes without adjustment. A single factor multiplying the forward-projected prediction, fitted to the 29 measured imaginary parts at 90 mT, is $1.01\pm0.05$.
The measured coefficients correlate with the sum at $r=0.91$, but with the temperature-gradient and curvature contributions alone at only $r=0.51$ and $-0.36$.
The correlation coefficient does not depend on scale, and no choice of strength for a single contribution therefore reproduces the measured pattern.
The two contributions are themselves correlated at $r=-0.98$, so the measured pattern is set by the small part of each that does not cancel.
For each triad, the degree of cancellation was quantified as the ratio of the smaller to the larger magnitude of the imaginary parts of the two forward-projected contributions.
These contributions are oppositely signed for all 29 retained triads, and the ratio ranges from 0.63 to 0.96.
\par
To assess how closely the interacting modes are locked to a common rotating structure, triad frequency matching and phase coherence were characterized by
\begin{equation}
    \delta_j=
    \frac{|\omega_{m_3}-\omega_{m_1}-\omega_{m_2}|}{\Gamma_j},
    \qquad b_j^2=
    \frac{|B_j|^2}
    {\langle|X_{m_3}|^2\rangle\langle|X_{m_1}X_{m_2}|^2\rangle}.
\end{equation}
Here, $\Gamma_j$ is the combined linewidth of the three participating spectral peaks, $\Gamma_j=(\Gamma_{m_3}^2+\Gamma_{m_2}^2+\Gamma_{m_1}^2)^{1/2}$.
The linewidths are added in quadrature because the mismatch $\omega_{m_3}-\omega_{m_1}-\omega_{m_2}$ combines three spectral lines with independent frequency spreads, whose spread is the quadrature sum of the individual ones.
A linear sum, appropriate for purely lifetime-broadened lines, would reduce each $\delta_j$ by at most a factor of $\sqrt3$ and leaves the ordering of the three field settings unchanged.
For each field setting, the bicoherence level expected from noise alone was evaluated at 95\% confidence following Refs.~\cite{elgar1988,elgar1989}, and both quantities are shown for the three field settings in Extended Data Fig.~\ref{fig:ex1}.

\bmhead{Data availability}
The coupling coefficients, transfer rates and resolution matrices underlying Fig.~\ref{fig:3} and Extended Data Fig.~\ref{fig:exisat} are available with the code via Zenodo at \url{https://doi.org/10.5281/zenodo.23657406}. The raw probe data and other data supporting the findings of this study are available from the corresponding authors upon reasonable request.

\backmatter

\bmhead{Code availability}
The code used for the quadratic system identification and for the calculation and comparison of the theoretical coupling coefficients is available via Zenodo at \url{https://doi.org/10.5281/zenodo.23657406}.

\bmhead{Acknowledgements}
The authors thank I. Niiya for valuable technical assistance during the experiments.
H.C. thanks the Plasma Laboratory for Advanced REsearch (PLARE) group at Seoul National University for their valuable support.
We used a large language model to assist in editing the manuscript and the analysis code.
This work was supported by JSPS KAKENHI (Grant Nos. 26K22312 and 24K06996) and in part by the Collaborative Research Program of the Research Institute for Applied Mechanics, Kyushu University.
This research was supported by R\&D Program of ``Optimal Basic Design of DEMO Fusion Reactor, CN2602-2'' through the Korea Institute of Fusion Energy (KFE) funded by the Government funds.

\bmhead{Author contributions}
H.C. conceived and designed the study.
H.C., D.D.M., T.Y. and C.M. performed the experiments, and D.D.M., T.Y. and C.M. operated PANTA and developed the probe arrays and data acquisition system.
H.C. implemented the system-identification analysis and wrote the analysis code, and H.C., G.C. and C.H. analysed the data.
H.C. and G.C. derived the extended drift-wave equation, and H.C. performed the numerical calculations of the coupling coefficients.
H.C., G.C., C.H., P.M., C.M. and Y.-S.N. interpreted the results.
H.C. wrote the original draft, and C.M. and Y.-S.N. revised it.
C.M. supervised the experiments, and Y.-S.N. supervised the project.
C.M. and Y.-S.N. acquired funding.
All authors discussed the results and commented on the manuscript.

\bmhead{Competing interests}
The authors declare no competing interests.

\bmhead{Correspondence}
Correspondence and requests for materials should be addressed to Chanho Moon or Yong-Su Na.

\setcounter{figure}{0}
\renewcommand{\figurename}{Extended Data Fig.}
\renewcommand{\theHfigure}{ED\arabic{figure}}
\begin{figure}
    \centering
    \includegraphics[width=1\linewidth]{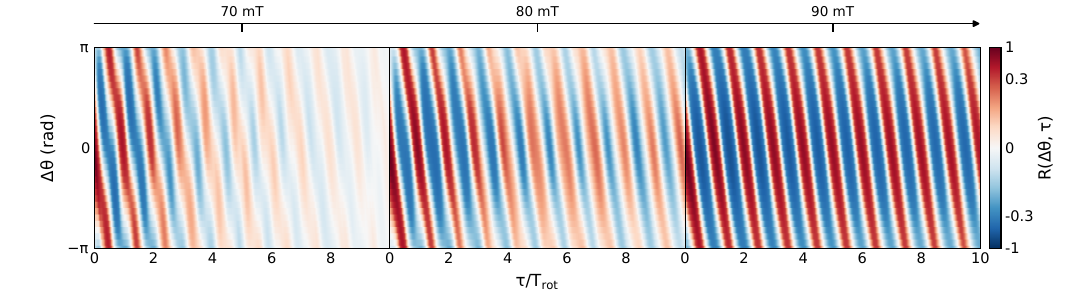}
    \caption{Azimuth--time correlation at the three field settings.
    Normalized azimuth--time correlation $R(\Delta\theta,\tau)$ measured with the azimuthal array at $r=4$ cm at 70, 80 and 90 mT (increasing along the arrow).
    The time lag is normalized by the corresponding rotation period $T_{\rm rot}$; the colour scale is symmetric logarithmic and linear for $|R|\le0.3$.}
    \label{fig:excorr}
\end{figure}
\begin{figure}
    \centering
    \includegraphics[width=0.49\linewidth]{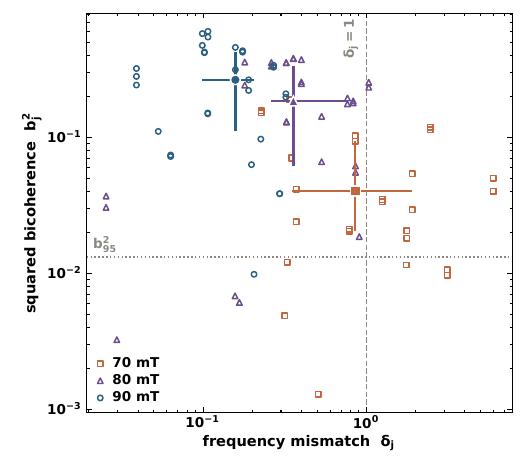}
    \caption{Triad frequency matching and phase coherence at the three field settings.
    Squared bicoherence $b_j^2$ versus triad frequency mismatch $\delta_j$, normalized by the combined linewidth of the three participating spectral peaks (Methods), for the 29 retained triads at 70 mT (squares), 80 mT (triangles) and 90 mT (circles).
    Open symbols denote individual triads. Filled symbols mark the median for each field setting, and bars span the central half of its triads.
    The vertical dashed line marks $\delta_j=1$, where the mismatch equals the linewidth, and the horizontal dotted line marks the bicoherence level $b_{95}^2$ expected from noise alone, taken as the largest of the three field settings (Methods).
    From 90 mT to 70 mT, the triads become less closely frequency-matched and less phase-coherent.}
    \label{fig:ex1}
\end{figure}
\begin{figure}
    \centering
    \includegraphics{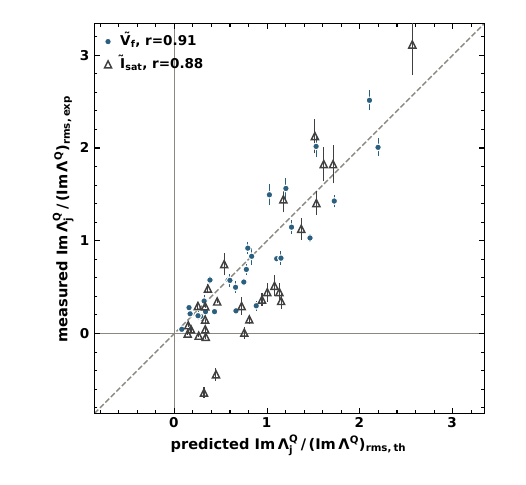}
    \caption{Coupling coefficients identified from floating-potential and ion-saturation-current signals at 90 mT.
    Imaginary parts of the coefficients identified from $e\widetilde V_{\rm f}/T_{e0}$ (filled circles) and from $\tilde I_{\rm sat}/I_{\rm sat0}$ (open triangles) versus the forward-projected theoretical coefficients for the 29 retained triads, with the resolution matrix of each signal applied separately.
    Measured and predicted coefficients are each normalized by their own root-mean-square value, so the comparison tests the triad dependence independently of the overall scale. The dashed line marks equal values and $r$ denotes the correlation coefficient.
    Error bars show the standard error over the 39 realizations of duration $16T_{\rm rot}$ in the stationary interval, enlarged by a factor of $\sqrt2$ to account for their 50\% overlap.}
    \label{fig:exisat}
\end{figure}
\end{document}